\documentclass[%
 reprint,superscriptaddress,groupedaddress
 amsmath,amssymb,
 aps]{revtex4-2}

\usepackage{graphicx}
\usepackage{dcolumn}
\usepackage{bm}
\usepackage{siunitx}
\usepackage{braket}
\usepackage{amsmath,amssymb,amsfonts}
\usepackage{hyperref}
\hypersetup{colorlinks=true,allcolors=blue}
\usepackage{todonotes}
\newcommand{\wrap}[1]{\parbox{.3\linewidth}{\vspace{1.5mm}#1\vspace{1mm}}}

\newcommand{\UIBK}{Institute f{\"u}r Experimentalphysik, Universit{\"a}t Innsbruck, Innsbruck, Austria}
\newcommand{\WWU}{Institut f{\"u}r Festk{\"o}rpertheorie, Universit{\"a}t M{\"u}nster, 48149 M{\"u}nster, Germany}
\newcommand{\TUdo}{Condensed Matter Theory, Department of Physics, TU Dortmund, 44221 Dortmund, Germany}
\newcommand{\Bayreuth}{Theoretische Physik III, Universit{\"a}t Bayreuth, 95440 Bayreuth, Germany}

\begin{document}

\preprint{arxiv}

\title{Dressed-state analysis of two-color excitation schemes}

\author{Thomas K. Bracht}%
    \affiliation{\WWU}
    \email{t.bracht@wwu.de}
    \affiliation{\TUdo}
\author{Tim Seidelmann}%
    \affiliation{\Bayreuth}
\author{Yusuf Karli}
    \affiliation{\UIBK}
\author{Florian Kappe}
    \affiliation{\UIBK}
\author{Vikas Remesh}%
    \affiliation{\UIBK}
\author{Gregor Weihs}%
    \affiliation{\UIBK}
\author{Vollrath Martin Axt}%
    \affiliation{\Bayreuth}
\author{Doris E. Reiter}%
    \affiliation{\TUdo}


\date{\today}

\begin{abstract}
To coherently control a few-level quantum emitter, typically pulses with an energy resonant to the transition energy are applied making use of the Rabi mechanism, while a single off-resonant pulse does not result in a population inversion. Surprisingly, a two-color excitation with a combination of two off-resonant pulses making use of the Swing-UP of quantum EmitteR population (SUPER) mechanism is able to invert the system. In this paper, we provide an in-depth analysis of the SUPER scheme within the dressed-state picture. We show that the SUPER mechanism can be understood as a driving of the transition between the dressed states. In the two-level system this allows us to derive analytic expressions for the pulse parameters yielding a population inversion. We extend our considerations to the three-level system, where a strong mixing between the bare states takes place. The insights gained from these results help in finding accessible regimes for further experimental realizations of the SUPER scheme and similar two-color excitation schemes. 
\end{abstract}

\maketitle

\section{Introduction}
Extensive efforts are put into finding excitation protocols for quantum emitters, which can be used for the deterministic and efficient generation of single photons \cite{Michler2000,Santori2001,He2013,wei2014deterministic,tonndorf15single,Somaschi2016,Ding2016,aharonovich2016,senellart2017high,chakraborty2019advances,rodt2020deterministically,zhai2022quantum,Schroder2011ultrabright,Schroder2016integrated,Leifgen2014evaluation,Englund2010,Beha2012,Fehler2019,Janitz2020,Schrinner2020,Proscia2020,Froech2020,grosso2017tunable} or entangled photon pairs \cite{vajner2022quantum,schimpf2021quantumcrypto,basso2021quantumkey,muller2014demand,akopian2006entangled,yin2017satellite,Huber2018semiconductor,pan2012multiphoton,stevenson2006semiconductor,young2006improved,hafenbrak2007triggered,muller2009creating,dousse2010ultrabright,stevensons2012indistuingishable,mueller2014ondemand,trotta2014highly,winik2017ondemand,huber2017highly,Bounouar2018generation,wang2019ondemand,liu2019solidstate,fognini2019dephasing,hopfmann2021maximally,cygorek2018comparison,seidelmann2019from,seidelmann2019phonon}, both being essential building blocks in quantum information technology. While well-established schemes like resonant excitation, leading to Rabi rotations \cite{stievater2001rabi,kamada2001exciton,ramsay2010review}, or adiabatic rapid passage (ARP) \cite{simon2011robust,wu2011population,kaldewey2017demonstrating,mathew2014subpico,debnath2012chirped} exist to excite a quantum emitter, the filtering to exclude resonant photons from the exciting laser essentially halves the photon output. There are several schemes available that circumvent this problem by allowing excitation using a detuned laser, like phonon assisted preparation \cite{Ardelt2014,bounouar2015phononassisted,quilter2015phononassisted,barth2016fast}, dichromatic excitation \cite{he2019coherently,koong2021coherent} or spectrally shaped ARP \cite{wilbur2022spectrally} or excitation via the biexciton \cite{sbresny2022stimulated}. The photon loss induced by filtering can also be reduced by embedding the emitter in a suited photonic structure \cite{uppu2020scalcable,wang2019towards,tomm2021bright}.

A radically different approach is the Swing-UP of quantum EmitteR population (SUPER) scheme \cite{bracht21swingup,shi2021two,bracht2022phonon, karli2022super} that exploits the coherence of the quantum system by using off-resonant pulse pairs. The SUPER scheme also circumvents the necessity for polarization filtering, without relying on incoherent or emitter-specific processes like phonon effects in a quantum dot, because the exciting laser frequency is energetically separated from the signal by several millielectronvolts.

In this paper, we analyze the mechanism of a multi-pulse scheme like SUPER in a dressed-state picture. The dressed states are the eigenstates of a coupled system-light Hamiltonian. Because the dressed-state description includes the system and its driving in a single picture, an in-depth understanding on how a specific scheme functions can be obtained
\cite{barth2016fast,lueker2012influence,lueker2017phonon}. We show how the dressed-state picture is applied to a multi-pulse scheme, where the pulses have different detunings with respect to the  transition energy of the quantum emitter. 
The analysis allows us to find analytic expressions for the pulse parameters that are needed for optimal excitation in the SUPER protocol. We analyze a two-level system as well as the three-level exciton-biexciton system, which is commonly found in semiconductor quantum dots.

\subsection{Dressed states}
\begin{figure}
    \centering
    \includegraphics{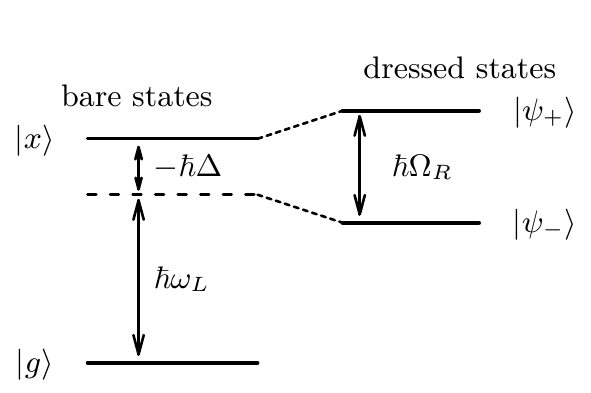}
    \caption{Two-level system with ground state $\ket{g}$ and excited state $\ket{x}$, showing laser energy and detuning. In the dressed state basis, the laser leads to an energy splitting with the Rabi energy.
    }
    \label{fig:level_scheme}
\end{figure}
We start with a brief summary of the concept of dressed states for a general two-level system with the transition energy $\hbar\omega_0$ between ground state $\ket{g}$ and an excited state $\ket{x}$, which is driven by an external source $\Omega(t)=\Omega_0(t) e^{-i\omega_L t}$ having a constant frequency $\omega_L$ and a time-dependent envelope $\Omega_0(t)$. This yields the standard Hamiltonian in the dipole- and rotating wave approximations (RWA)
\begin{equation}
    H_{\text{TLS}} = \hbar\omega_0\ket{x}\bra{x}-\frac{\hbar}{2}\left( \Omega(t) \ket{x}\bra{g}+h.c.\right)\,. \label{eq:TLS}
\end{equation}
Defining the detuning $\Delta=\omega_L-\omega_0$, the Hamiltonian can be rewritten in a rotating frame as
\begin{equation} \label{eq:TLS_RF}
    H_{\text{TLS,RF}} = -\hbar \Delta \ket{x}\bra{x}-\frac{\hbar}{2}\left(\Omega_0(t)\ket{x}\bra{g}+h.c.\right) \,.
\end{equation}
The term \textit{dressed states} refers to the bare states $\ket{g},\ket{x}$ that are \textit{dressed} with the light field. They are calculated by diagonalizing the bare state Hamiltonian and the electron-light interaction Hamiltonian. For constant fields, this is commonly done \cite{tannor2007introduction}, while for time-dependent fields the transformation is performed for each point in time.

Diagonalizing the Hamiltonian in Eq.~\eqref{eq:TLS_RF} results in 
\begin{equation} 
    \Tilde{H}_{\text{TLS}} = E_{+}\ket{\psi_+}\bra{\psi_+} + E_{-}\ket{\psi_-}\bra{\psi_-},
\end{equation}
with the dressed states $\ket{\psi_{\pm}}$ and the dressed-state energies $E_{\pm}$
\begin{equation}
    E_{\pm} = \frac{\hbar}{2}\left(-\Delta \pm \sqrt{\Omega_0^2(t)+\Delta^2}\right).
    \label{eq:dressed_state_energies}
\end{equation} 
with their difference being the Rabi splitting 
\begin{equation}
   E_+ - E_{-} = \hbar \Omega_R(t) =\hbar \sqrt{\Omega_0^2(t)+\Delta^2}.
   \label{eq:TLS_Rabi}
\end{equation}
The splitting of the dressed-state energies in relation to the bare-state energies is schematically shown in Fig.~\ref{fig:level_scheme}.

The dressed states can be written as superpositions of the bare states
\begin{align}
    \begin{split}
        \ket{\psi_+}&= c(t) \ket{x} - \Tilde{c}(t)\ket{g},\\
        \ket{\psi_-}&= \Tilde{c}(t)\ket{x} + c(t)\ket{g}.
    \end{split}
    \label{eq:dressed_states}
\end{align}
with the coefficients 
\begin{align}
\begin{split}
    c(t) &= \frac{\Omega_R(t)-\Delta}{\sqrt{\Omega_0^2(t) + (\Omega_R(t) - \Delta)^2}},\\
    \Tilde{c}(t)  &= \sqrt{1-c^2(t)}.
\end{split}\label{eq:dressed_factors}
\end{align}
Other than in the bare state picture, where $\ket{g},\ket{x}$ are time-independent, in the dressed state picture the time-dependence of the driving is accounted for by the state mixing and the dressed-state energies $E_\pm$.

\subsection{Idea of SUPER}
The concept behind the SUPER scheme, as it was proposed in Ref.~\cite{bracht21swingup}, relies on the usage of two detuned pulses to drive a transition in a few-level system. The two laser frequencies lead to a beating effect, which gives rise to changes in the excited state population. Assuming that the system is initially in the ground state and the pulse parameters are tuned correctly, the system experiences a swing-up effect and ends up in the excited state. 
The multi-color pulse pair is described by
\begin{equation}
    \Omega_X(t) = \Omega_1(t)e^{-i\omega_1 t } + \Omega_2(t)e^{-i\omega_2 t}\,.
    \label{eq:pulses} 
\end{equation}
Here, $\omega_j$ are the central frequencies of the pulses and $\Omega_j(t)$ are the (real) pulse envelopes. Note that the pulses do not have to fulfill a strict phase relationship, as shown in Ref.~\cite{bracht21swingup}. Phonons are known to result in decoherence effects for state preparation schemes in quantum dots \cite{luker2019review}, which here have been shown to have rather negligible effects \cite{bracht2022phonon}. For clarity, we therefore neglect dissipation in this paper. Essential properties of the laser pulses are their respective detunings to the transition frequency of the quantum emitter $\omega_0$ defined as
\begin{equation}
    \Delta_j = \omega_j - \omega_0.
\end{equation}
In the following, we will refer to the two pulses by the `first' pulse with detuning $\Delta_1$ and the `second' pulse with detuning $\Delta_2$. This naming convention refers to the idea that the parameters for the first pulse are fixed and determine the parameters of the second pulse. Hence, the naming shall not necessarily indicate
a temporal order. We remind the reader that the two pulses should overlap in time for SUPER to work. Most importantly, it was shown in Ref.~\cite{bracht21swingup} that the frequency difference of the pulses should coincide with the Rabi frequency induced by the first pulse. This means, the pulse energies can not be chosen arbitrarily, but have a fixed relation to each other. In the following, we show how this relation can be extracted from the dressed state picture. 

%
\section{Dressed state analysis of a two-level system}
We begin the study of the influence of the multi-color pulses on a two-level system by replacing the single laser pulse in Eq.~\eqref{eq:TLS} with the multi-color pulse as defined in Eq.~\eqref{eq:pulses}, within the Hamiltonian 
\begin{equation}
    H = \hbar\omega_0\ket{x}\bra{x}-\frac{\hbar}{2}\left(\Omega_X(t)\ket{x}\bra{g}+h.c.\right). \label{eq:hamiltonian_2ls}
\end{equation}
Because the Hamiltonian now contains two laser frequencies, the transformation into the rotating frame is somewhat ambiguous, as several frequencies can be chosen. We decide to transform the Hamiltonian in Eq.~\eqref{eq:hamiltonian_2ls} to a reference frame rotating with frequency $\omega_1$ of the first pulse, yielding
\begin{align}
\begin{split}
        H_{RF} = &-\hbar\Delta_1\ket{x}\bra{x}-\frac{\hbar}{2}\Omega_1(t)\left(\ket{x}\bra{g}+\ket{g}\bra{x}\right)\\
        & -\frac{\hbar}{2}\Omega_2(t)\left(e^{i\omega_{\Delta}t}\ket{x}\bra{g}+e^{-i\omega_{\Delta}t}\ket{g}\bra{x}\right)\\
        = & H_0 + H_{\Omega_1} + H_{\Omega_2} \,.
\end{split}\label{eq:hamiltonian_rf}
\end{align}
After the rotating frame transformation, $H_0+H_{\Omega_1}$ is purely real, while $H_{\Omega_2}$ still includes terms oscillating with the frequency difference between the two pulses, which we define as
\begin{equation}
    \omega_{\Delta}=\omega_1-\omega_2 = \Delta_1-\Delta_2.\label{eq:omega_delta}
\end{equation}
Note that $\omega_\Delta$ is a difference of two frequencies and can take on negative values. 

%
\subsection{Dressed states including the first pulse}
To analyze the energetic splitting induced by the first pulse, a transformation to a dressed state basis is performed, including only the first pulse and neglecting the second pulse for now. In some sense, these are partially dressed states. In other words, we diagonalize $H_0 + H_{\Omega_1}$ in Eq.~\eqref{eq:hamiltonian_rf}, leading to the energies and states as given in Eq.~\eqref{eq:dressed_state_energies} and Eq.~\eqref{eq:dressed_states}.

For illustrative purposes, let us consider a system, where these partially dressed states are approximately time-independent. We achieve this by assuming a rectangular shaped pulse, though such pulses are challenging to realize with state-of-the-art laser systems \cite{neumann2021optical}. To avoid instantaneous switching of the pulse that leads to many additional components in the laser spectrum and to allow for dressing and undressing processes \cite{barth2016fast}, we consider a switch on/off process of finite duration. 
This is modeled using two Sigmoid-type functions to assure smooth edges:
\begin{equation}
    \Omega_1(t) = \frac{\Omega_1^{\text{rect}}}{(1+e^{-\kappa(\tau/2+t)})(1+e^{-\kappa(\tau/2-t)})}\label{eq:rectangular_pulse}
\end{equation}
This pulse has an approximate plateau duration of $\tau$, while $\kappa$ determines the duration of the pulse edges. $\Omega_1^{\text{rect}}$ is a measure for the amplitude of the pulse during its \textit{on} state. 
During the plateau region of the pulse, the dressed states are essentially constant, such that we do not have to account for the time dependence of the dressed state quantities. Noticeable changes only happen during the dressing and undressing process. 

As parameters we choose  a detuning of $\hbar\Delta_1=\SI{-5}{meV}$ and an amplitude of $\hbar \Omega_1^{\text{rect}} = \SI{4}{meV}$. These parameters are typical when dealing with the optical excitation of quantum dots and the excited-state lifetime is about an order of magnitude larger than the pulse duration. Figure~\ref{fig:dressed_states_rectangle_diff}(a) shows the dressed-state energies for this rectangular pulse shape with $\tau=\SI{40}{ps}$ and $\kappa=\SI{1}{ps^{-1}}$. Before and after the pulse, the energies are split by the detuning $\hbar\Delta_1$ of the laser due to the frame of reference. During the pulse, the branches shift apart by a constant splitting of $\SI{6.4}{meV}$, as described by Eq.~\eqref{eq:TLS_Rabi}. Before and after the pulse, the upper dressed state $\ket{\psi_+}$ can be identified with the excited state $\ket{x}$, while the lower dressed state $\ket{\psi_-}$ can be identified with the ground state $\ket{g}$. 
%
%
\subsection{Transformation of the second pulse}
In the next step, we study the influence of the second pulse, interacting with the system at the same time as the first pulse. Correspondingly, the interaction Hamiltonian $\Tilde{H}_{\Omega_2}$ is transformed to the dressed state picture of the first pulse. The inverse transformation to Eq.~\eqref{eq:dressed_states} leads to the transformed Hamiltonian
\begin{align}
    \begin{split}
    \Tilde{H}_{\Omega_2} =& \hbar\Omega_2(t)c\Tilde{c}\,\cos(\omega_{\Delta}t)(\ket{\psi_+}\bra{\psi_+}-\ket{\psi_-}\bra{\psi_-})\\
    &-\frac{\hbar}{2}\Omega_2(t)\left[\left(c^2e^{i\omega_{\Delta}t} - \Tilde{c}^2e^{-i\omega_{\Delta}t}\right)\ket{\psi_+}\bra{\psi_-} + h.c.\right].
    \end{split}\label{eq:interaction_dressed}
\end{align}
On one hand, the Hamiltonian $\tilde{H}_{\Omega_2}$ leads to additional energy shifts due to the $\ket{\psi_\pm}\bra{\psi_\pm}$ terms. These energy shifts cause an oscillation of the dressed-state energies $E_{\pm}$,
\begin{equation}
    E_{\pm} \rightarrow E_{\pm,2} = E_{\pm} \pm \hbar\Omega_2(t) c \tilde{c} \cos(\omega_\Delta t).\label{eq:energy_new}
\end{equation}
On the other hand, $\tilde{H}_{\Omega_2}$ also leads to transitions between the dressed states, mediated by the $\ket{\psi_\pm}\bra{\psi_\mp}$ terms. The term $\propto c^2 e^{i\omega_{\Delta}t}$ as well as the term $\propto \Tilde{c}^2 e^{-i\omega_{\Delta}t}$ in Eq.~\eqref{eq:interaction_dressed} drives these transitions. In analogy to the usual two-level system, we arrive at 
\begin{align}
    \begin{split}
    \Tilde{H} =& E_{+,2}\ket{\psi_+}\bra{\psi_+} + E_{-,2}\ket{\psi_-}\bra{\psi_-}\\
    &-\frac{\hbar}{2}\Omega_2(t)\left[\left(c^2e^{i\omega_{\Delta}t} + \Tilde{c}^2e^{-i\omega_{\Delta}t+i\pi}\right)\ket{\psi_+}\bra{\psi_-} + h.c.\right].
    \end{split}\label{eq:interaction_dressed_new}
\end{align}
This resembles the Hamiltonian of a two-level system, consisting of the two states $\ket{\psi_+},\ket{\psi_-}$, driven by the second laser pulse, while the driving $\Omega_2(t)$ is modified by the dressed state coefficients. An interesting aspect is, that it contains both the positive and the negative frequency component $e^{\pm i\omega_{\Delta}t}$. This is in analogy to a Hamiltonian where the RWA has not been performed. We remind, that in the RWA for the raising operator $\ket{\psi_+}\bra{\psi_-}$ only terms with $e^{-i|\omega|t}$ are considered. However, because $\omega_\Delta$ can take positive and negative values, both terms are important in Eq.~\eqref{eq:interaction_dressed_new}. We distinguish two cases: (i) driving based on $e^{-i\omega_{\Delta}t+i\pi}$, where the additional phase of $\pi$ will be neglected in the following, (ii) driving based on $e^{+i\omega_{\Delta}t}$.

For now, we focus on case (i). Similar to a standard two-level system, the raising operation $\propto e^{-i\omega_\Delta t}$ is resonant, if the frequency $\omega_\Delta>0$ corresponds to the energy difference of the two levels, that is
\begin{equation}
    \hbar\omega_\Delta = E_{+,2}-E_{-,2} = E_+-E_-+2\hbar\Omega_2(t)c\tilde{c}\cos(\omega_\Delta t)
\end{equation}
However, this energy difference is time-dependent and depends on $\omega_\Delta$ itself. 
In order to find the resonance condition for the second pulse, we take the mean energy difference, neglecting the fast oscillations of the dressed-state energies. Using Eq.~\eqref{eq:omega_delta}, this gives us the condition for the detuning of the second pulse with
\begin{align}
    \begin{split}
        \omega_\Delta &= \left(E_+ - E_-\right)/\hbar = \Omega_R = \sqrt{\left(\Omega_1^{\text{rect}}\right)^2+\Delta_1^2}\\
        \Rightarrow\Delta_2 &= \Delta_1 -\Omega_R.\label{eq:energy2_below}
    \end{split}
\end{align}
In this configuration, the second laser pulse is always energetically below the first one.\\

\subsection{Driving the dressed two-level system}
Keeping these results in mind, we now consider the scenario shown in Fig.~\ref{fig:dressed_states_rectangle_diff}(a) and search for a second pulse to excite the system. 
For the second pulse, we use a Gaussian pulse envelope given by
\begin{equation}
    \Omega_2(t) = \frac{\alpha_2}{\sqrt{2\pi}\sigma_2}e^{-\frac{t^2}{2\sigma_2^2}},
\end{equation}
with pulse area $\alpha_2$ and duration $\sigma_2$.
The duration of the second pulse has to be chosen such that it lies completely inside the plateau region of the first pulse. 

In analogy with the bare-state two-level system, we now search for a pulse, that results in an inversion between the dressed states, i.e., which transfers the system completely from its initial state $\ket{\psi_-}$ to $\ket{\psi_+}$. In other words a $\pi-$rotation has to be performed in the dressed state picture. Looking at the Hamiltonian in Eq.~\eqref{eq:interaction_dressed_new}, the pulse area (and Rabi frequency) of the second pulse is effectively reduced by the factor $\tilde{c}^2$. 
Due to this renormalization of the pulse area, $\alpha_2$ has to be scaled by this factor such that $\tilde{c}^2\alpha_2=\pi$, yielding the condition 
\begin{equation} 
    \alpha_2 = \frac{\pi}{\tilde{c}^2}
    \label{eq:pulsearea}
\end{equation}

Here, we use $\sigma_2=\SI{4}{ps}$.
We choose the detuning of the second pulse to be smaller than the first one, i.e., using the raising operation $\propto e^{-i\omega_\Delta t}$ with $\omega_{\Delta}>0$.

\subsubsection{Negative $\Delta_1$}
As the first pulse has a negative detuning of $\hbar\Delta_1=\SI{-5}{meV}$, both exciting lasers are below the excited state transition. From Eq.~\eqref{eq:energy2_below} using $\hbar\Omega_R=\SI{6.4}{meV}$, we calculate a detuning of $\hbar\Delta_2 = \SI{-11.4}{meV}$ for the second pulse. The necessary pulse area follows from Eq.~\eqref{eq:pulsearea} and evaluates to $\alpha_2 \approx 9.13\pi$. We stress that the first pulse completely determines the properties of the second pulse needed to perform a rotation in the dressed state basis.

\begin{figure}[t]
    \centering
    \includegraphics{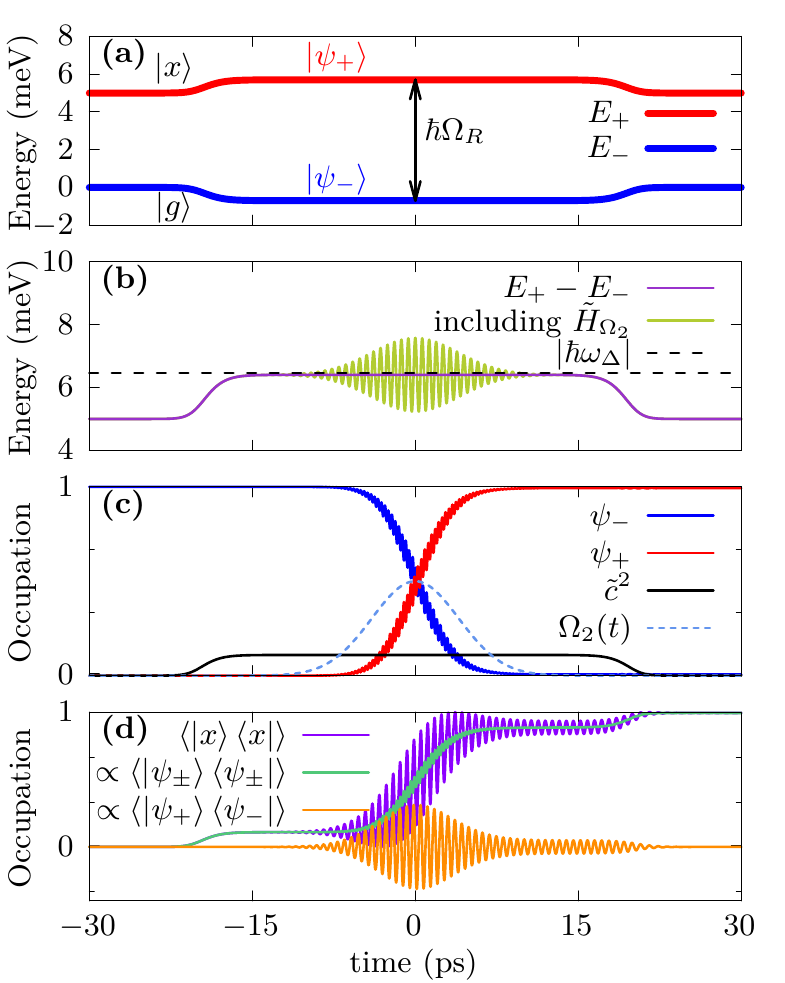}
    \caption{
    \label{fig:dressed_states_rectangle_diff}
    Time evolution of several quantities: (a) Energies of the dressed states for a rectangular pulse with smoothened edges with $\hbar\Delta_1=\SI{-5}{meV},\tau=\SI{40}{ps},\kappa=\SI{1}{ps^{-1}},\hbar\Omega_0=\SI{4}{meV}$.
    (b) Energy difference of upper and lower dressed state 
    excluding (violet) and including (green) the influence of the second pulse. The dashed black line indicates the detuning of the second pulse, which in the rotating frame is $|\hbar\omega_\Delta|$. 
    (c) Dressed-state occupation under the second pulse with $\sigma_2=\SI{4}{ps},\alpha_2=9.13\pi,\hbar\Delta_2=\SI{-11.46}{meV}$ where the second pulse is indicated by the dashed line. 
    (d) Occupation of the excited state in the bare state picture as well as contributions of the dressed states.
    }
\end{figure}

Figure~\ref{fig:dressed_states_rectangle_diff} shows the effect of this second pulse on the system. In Fig.~\ref{fig:dressed_states_rectangle_diff}(b), the energy difference of the dressed states is shown in violet, while the green line includes the contributions of the second pulse. This leads to the additional oscillatory splitting of the dressed states described in Eq.~\eqref{eq:energy_new}. The black dashed line indicates the energy of the second laser. 

For $\Delta_2$, we determined numerically that an additional shift of about $\SI{-0.06}{meV}$ is needed to reach optimal excitation, i.e., $\hbar\Delta_2 = \SI{-11.46}{meV}$, which we attribute to perturbations due to the strong second pulse. In Fig.~\ref{fig:dressed_states_rectangle_diff}(c), the time evolution of the dressed-state occupation is shown alongside the coupling $\tilde{c}^2$ and the second laser envelope (normalized). The dressed-state occupation performs a Rabi-like oscillation from $\ket{\psi_{-}}$ to $\ket{\psi_{+}}$. During the rise of the dressed-state occupation, the occupation displays an additional low-amplitude high-frequency oscillation. The latter can be associated with the additional oscillations in the energy difference $E_{+,2}-E_{-,2}$ [see panel (b)].

Figure~\ref{fig:dressed_states_rectangle_diff}(d) shows the time evolution of the excited state population in purple. As intended, the occupation inverts during the action of both pulses from the ground to its excited state. The main rise is during the action of the second pulse, where also in the dressed-state picture, the second pulse mimicks a $\pi$-pulse that inverts the states. In addition, at the smoothened edges of the first pulse a rise of the excited state occupation of the magnitude of the state mixing of $\tilde{c}^2=0.11$ takes place. This adiabatic dressing and undressing effect is needed to fully invert the system, as has already been discussed for phonon-assisted state preparation schemes for exciton in quantum dots \cite{barth2016fast}. Hence, better results are obtained for smooth pulses like Eq.~\eqref{eq:rectangular_pulse} in contrast to sharp rectangular ones. 

The dressed-state picture can also be used for an even deeper analysis by separating
the fast, high amplitude oscillations in the bare states from the mean rise of the occupation.  
Expressing the excited-state occupation in terms of the dressed states reads
\begin{align}
    \begin{split}
        \braket{\ket{x}\bra{x}} =& c^2\braket{\ket{\psi_{+}}\bra{\psi_{+}}} + \tilde{c}^2\braket{\ket{\psi_{-}}\bra{\psi_{-}}}\\
        &+2c\tilde{c}\,\mathrm{Re}(\braket{\ket{\psi_{+}}\bra{\psi_{-}}}).
    \end{split}\label{eq:direct_indirect_contribution}
\end{align}
Here, the first line of Eq.~\eqref{eq:direct_indirect_contribution} can be interpreted as the contribution of the dressed state population to the bare state occupations due to mixing.
The part in the second line depends on the term $\braket{\ket{\psi_{+}}\bra{\psi_{-}}}$, which represents the coherence between the dressed states. This coherence builds up with the second pulse and, similar to the coherence $\braket{\ket{g}\bra{x}}$ between the bare states, gives rise to transitions between the dressed states. However, additionally, it leads to a contribution to the bare state occupation. A sizable influence of this term is the fingerprint of a non-adiabatic time-evolution.

Both parts contributing to $\braket{\ket{x}\bra{x}}$ are shown in Fig.~\ref{fig:dressed_states_rectangle_diff}(d). Comparing the two different contributions (green for population contribution and orange for the coherence contribution) to the exciton occupation $\braket{\ket{x}\bra{x}}$, we see that the mean behavior of the exciton occupation is determined by the population contribution. This holds in particular for the rise during the second pulse and the adiabatic dressing and undressing effects at the smoothened edges of the pulse. The coherence contribution starts during the second pulse and features a large-amplitude oscillation, which is also clearly visible in the excited state occupation. 

\begin{figure}[t]
    \centering
    \includegraphics{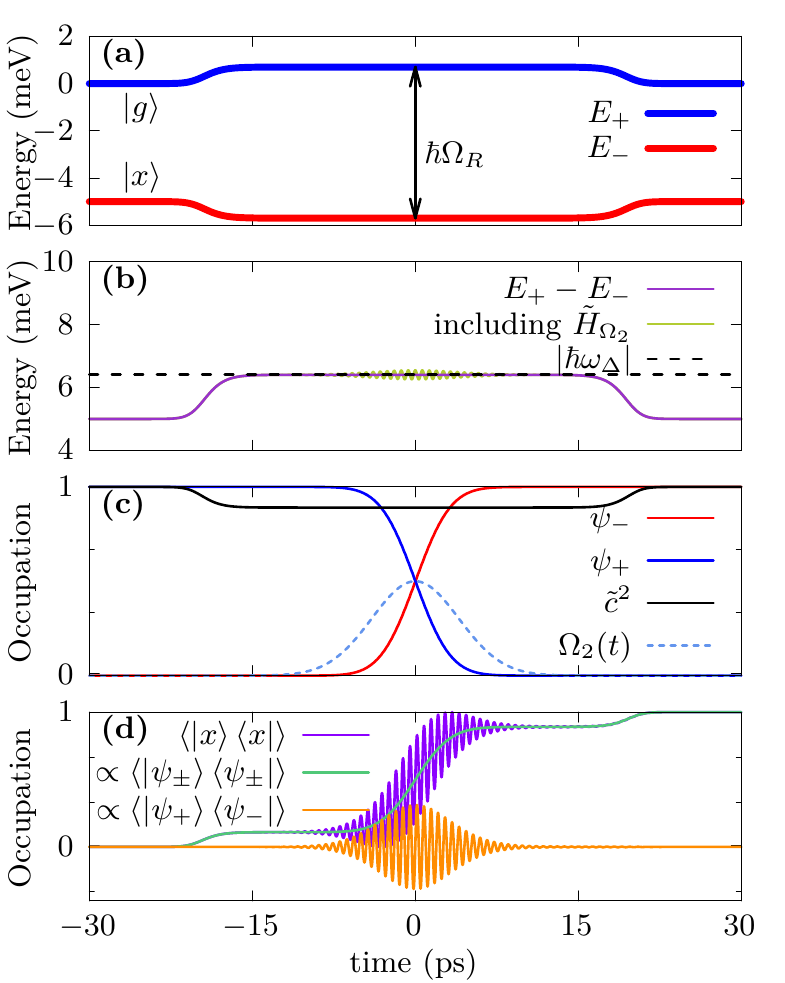}
    \caption{Same as Fig.~\ref{fig:dressed_states_rectangle_diff}, but for a positive detuning of $\hbar\Delta_1=\SI{+5}{meV}$, which yields as parameters for the second pulse $\sigma_2=\SI{4}{ps},\alpha_2=1.12\pi,\hbar\Delta_2=\SI{-1.41}{meV}$.
    }
    \label{fig:dressed_states_rectangle_positive}
\end{figure}
\subsubsection{Positive $\Delta_1$}
\begin{table}[]
    \centering
    \begin{tabular}{c|c|c}
        \wrap{$\hbar\Delta_2$ for} & $\hbar\Delta_1 =\SI{-5}{meV}$ & $\Delta_1 =\SI{5}{meV}$\\
        \hline
        $\omega_{\Delta} > 0$ & \wrap{$\SI{-11.46}{meV}$\newline (Fig.~\ref{fig:dressed_states_rectangle_diff})} & \wrap{$\SI{-1.41}{meV}$ \newline (Fig.~\ref{fig:dressed_states_rectangle_positive})} \\
        \hline
        $\omega_{\Delta} < 0$ & \wrap{$\SI{1.41}{meV}$} & \wrap{$\SI{11.46}{meV}$}
    \end{tabular}
    \caption{Detunings used for Figs.~\ref{fig:dressed_states_rectangle_diff} and \ref{fig:dressed_states_rectangle_positive}, also showing the combinations of detunings, when the second possible raising operation in Eq.~\eqref{eq:interaction_dressed_new} is used. We remind that the explicit detunings depend on the pulse strength.} 
    \label{tab:my_label}
\end{table}
So far, we focused on the case where both pulses are tuned below the excited state energy. However, Eq.~\eqref{eq:energy2_below} also holds for positive $\Delta_1$ (and still $\omega_{\Delta}>0$). Choosing the same shape of the first pulse as before, but using $\hbar\Delta_1=\SI{+5}{meV}$, leads to the dressed-state energies shown in Fig.~\ref{fig:dressed_states_rectangle_positive}(a). Now, before and after the pulse, the upper dressed state corresponds to the ground state and the lower dressed state corresponds to the excited state. This means that $c$ and $\tilde{c}$ change roles, i.e., now $\tilde{c}^2=0.89$. With these parameters and a splitting of the dressed states of $\hbar\Omega_R=\SI{6.4}{meV}$, according to Eq.~\eqref{eq:energy2_below} the energy of the second pulse evaluates to $\hbar\Delta_2=\SI{-1.413}{meV}$, which also agrees with the numerically found optimal parameters. The necessary pulse area evaluates to only $\alpha_2 =1.12\pi$, as the coupling coefficient $\tilde{c}^2$ is large [cf., Fig.~\ref{fig:dressed_states_rectangle_positive}(c)]. Energetically, this also leads to a lower impact of the second pulse on the dressed-state energies, such that it is resonant with the dressed-state transition for most of the time, even during the time it leads to an additional shift [see green line in Fig.~\ref{fig:dressed_states_rectangle_positive}(b)]. Figure \ref{fig:dressed_states_rectangle_positive}(d) shows the exciton population and the dressed-state contributions, where both the dressed-state occupation and the coherences contribute to the behaviour of the bare state occupation. 
Again, the dressed state picture provides a very useful analytic expression yielding parameters that lead to a complete occupation of the excited state.


\subsubsection{Negative $\omega_\Delta$}
We remind the reader that Eq.~\eqref{eq:interaction_dressed_new} includes two possible cases for $\omega_\Delta$ that lead to transitions. The second possible raising operation, $\propto e^{i\omega_\Delta t}$, is resonantly driven if we account for the opposite sign, i.e., the energetic condition changes to $- \hbar\omega_\Delta = E_{+,2}-E_{-,2}$. This is in agreement with $\omega_\Delta$ being able to take negative values and leads to a symmetry in the possible detunings, which is shown in Tab.~\ref{tab:my_label}. The signs of both detunings can be switched at the same time, under which the dynamics stay the same, except for a constant phase. Note that possible environmental effects, such as electron-phonon interaction in quantum dots, are not symmetric with respect to the detuning. 

\subsection{Gaussian Pulses}
\begin{figure}[t]
    \centering
    \includegraphics{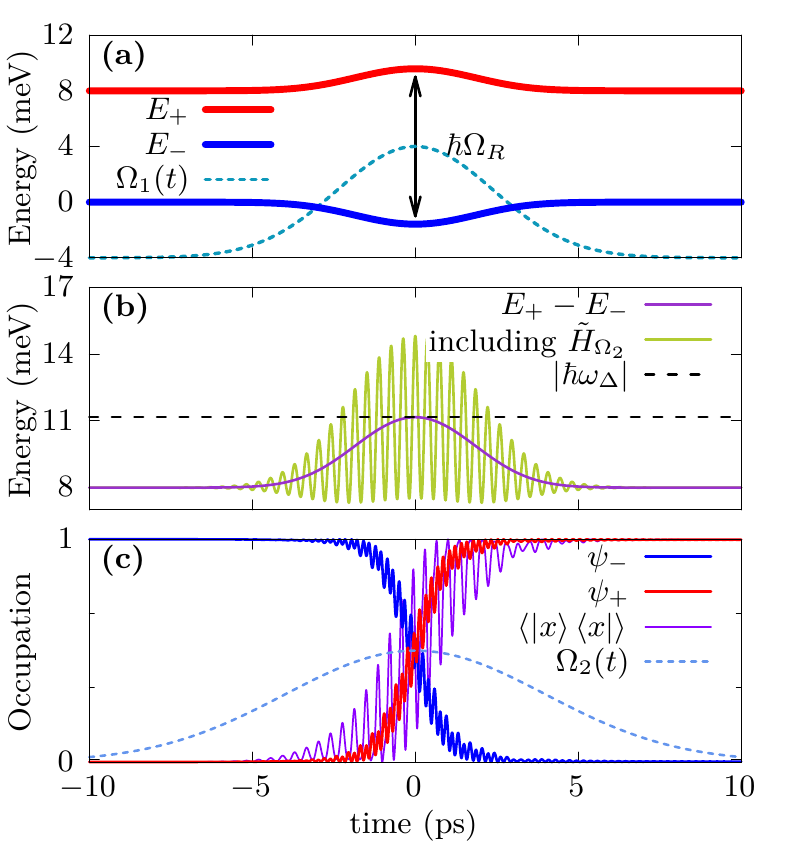}
    \caption{(a) Dressed-state energies $E_{\pm}$ induced by a single off-resonant Gaussian pulse, $\sigma_1=\SI{2.4}{ps},\hbar\Delta_1=\SI{-8}{meV}, \alpha_1=22.65\pi$, as indicated by the dashed line.  
    (b) Energy difference of the dressed states, (violet) without and (green) including the interaction Hamiltonian $\tilde{H}_{\Omega_2}$ (second pulse parameters: $\sigma_2=\SI{3.04}{ps},\hbar\Delta_2=\SI{-19.163}{meV}, \alpha_2=19.29\pi)$. The dashed black line depicts the energy of the second laser 
    (c) Dressed state and excited state occupation. The dashed blue line shows the shape of the second pulse.}
    \label{fig:dressed_state_diff}
\end{figure}%

In the proposal \cite{bracht21swingup} as well as in the experimental realisation \cite{karli2022super} two Gaussian pulses were used, such that also for the first pulse we now take a Gaussian envelope
\begin{equation}
    \Omega_1(t) = \frac{\alpha_1}{\sqrt{2\pi}\sigma_1}e^{-\frac{t^2}{2\sigma_1^2}} \,.
\end{equation}
with the parameter set of Ref.~\cite{bracht21swingup} with  $\sigma_1=\SI{2.4}{ps},\hbar\Delta_1=\SI{-8}{meV}, \alpha_1=22.65\pi$. The resulting dressed-state energies are shown in Fig.~\ref{fig:dressed_state_diff}(a). During the pulse, the branches split by up to $\SI{11.163}{meV}$. Using this maximum splitting in Eq.~\eqref{eq:energy2_below} leads to $\hbar\Delta_2=\SI{-19.163}{meV}$, the value also used in the original paper.

Here, at the maximum of the first pulse, $\tilde{c}^2=0.145$. But due to the continuous change in both pulse envelopes, the condition for the pulse area $\alpha_2$ as in Eq.~\eqref{eq:pulsearea} can not be applied.

The energy difference of the dressed states is displayed in Fig.~\ref{fig:dressed_state_diff}(b). The violet line depicts the energy difference of the two dressed states without the interaction Hamiltonian $\tilde{H}_{\Omega_2}$. When the additional energy shifts due to $\tilde{H}_{\Omega_2}$ are included, the oscillations according to Eq.~\eqref{eq:interaction_dressed} can be observed, shown as the green line. The dashed black line, marking the energy of the second laser in this rotating frame, crosses the green line several times. As soon as the energy matches, the dressed-state occupation starts to switch significantly from $\ket{\psi_-}$ to $\ket{\psi_+}$, as visible in Fig.~\ref{fig:dressed_state_diff}(c). The oscillating energy difference again leads to an oscillation in the dressed-state occupation as well. Although the second pulse is longer than the first one, the dynamics only happens in the time interval in which the energy-matching condition is fulfilled, that means, where the first pulse splits the dressed states. This is in accordance with the observation, that a single, highly detuned pulse does not lead to transitions in a simple two-level system, but the pulse pair is able to switch the occupation from the ground to the excited state.\\

%
%
\section{Three-level system}
Quantum dots can be used to generate entangled photons via the biexciton-exciton cascade. For this, a deterministic and high-fidelity preparation of the biexciton state $\ket{xx}$ is required. Therefore, we here analyze within the dressed state picture, how this can be achieved with a multi-pulse scheme like SUPER. 

Assuming linear polarization of the laser, the Hamiltonian including the biexciton can be written as a three-level system by adding the biexciton state $\ket{xx}$
\begin{align}
\begin{split}
    H = &\hbar\omega_0\ket{x}\bra{x} +\hbar(2\omega_0-\Delta_B)\ket{xx}\bra{xx} \\
&- \frac{\hbar}{2}\left[\Omega_X(t)\left(\ket{x}\bra{g}+\ket{xx}\bra{x}\right)+h.c.\right]. 
\end{split}\label{eq:hamiltonian_cl}
\end{align}
The energy of the biexciton is decreased by the binding energy $\hbar\Delta_B$ compared to double the exciton energy $\hbar\omega_0$. We set $\hbar\Delta_B=\SI{4}{meV}$, which is typical for quantum dots used in recent experiments \cite{karli2022super}.\\
%
Following the procedure from the two-level system, a transformation to a reference frame rotating with the frequency of the first laser leads to
\begin{align}
\begin{split}
    H_{RF} = &-\hbar\Delta_1\ket{x}\bra{x} -\hbar(2\Delta_1+\Delta_B)\ket{xx}\bra{xx} \\
  &-\frac{\hbar}{2}\Omega_1(t)\left[\left(\ket{x}\bra{g}+\ket{xx}\bra{x}\right)+h.c.\right]\\
  &-\frac{\hbar}{2}\Omega_2(t)\left[e^{i \omega_{\Delta}t}\left(\ket{x}\bra{g}+\ket{xx}\bra{x}\right)+h.c.\right]\\
  =& H_0 + H_{\Omega_1}+ H_{\Omega_2} \,.
\end{split}
\end{align}
\subsection{Effect of the first pulse}
Again, the Hamiltonian $H_0+H_{\Omega_1}$ is diagonalized to arrive at the dressed state picture, now consisting of the three dressed states
\begin{equation}
    \ket{\psi_j(t)}=\sum_{k\in\{g,x,xx\}} a_{j,k}(t)\ket{k}, \quad j=1,2,3.
\end{equation}
at the energies $E_j$. The coefficients $a_{j,k}(t)$ of the bare states are determined numerically.
\begin{figure}[t]
    \centering
    \includegraphics{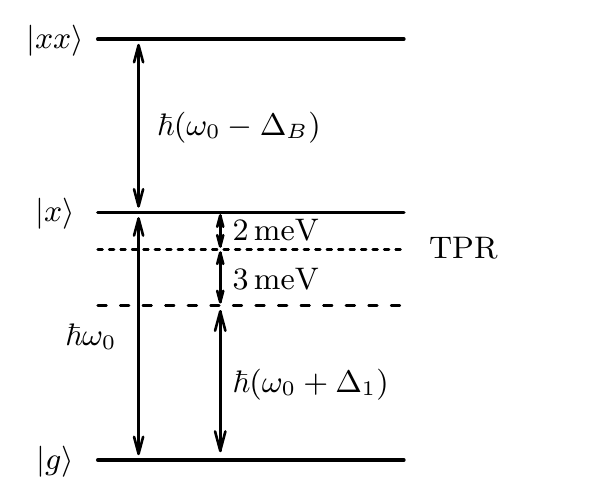}
    \caption{Three-level system with ground state $\ket{g}$, exciton state $\ket{x}$ and biexciton state $\ket{xx}$. The detuning of the first laser is tuned $\SI{3}{meV}$ below the two-photon resonance (TPR), which is located at $\SI{-2}{meV}$. 
    }
    \label{fig:scheme_3ls}
\end{figure}
For a first laser pulse with a detuning of $\hbar\Delta_1=\SI{-5}{meV}$ to the exciton transition, the laser is red-detuned to the two-photon resonance by $\SI{-3}{meV}$, like shown schematically in Fig.~\ref{fig:scheme_3ls}. For a pulse with $\alpha_1=27\pi$ and $\sigma_1=\SI{3}{ps}$, the dressed-state energies are shown in Fig.~\ref{fig:dressed_states_biexciton_coupling}(a). The three dressed states can be identified with the ground, exciton and biexciton state before and after the pulse. During the pulse, the energy branches bent in such a way, that there is a splitting of $\SI{12.31}{meV}$ between $E_3$ and $E_1$. The middle branch bends downward, yielding an energy separation of $\SI{5.88}{meV}$ between $E_2$ and $E_1$. The color of the branches in Fig.~\ref{fig:dressed_states_biexciton_coupling}(a) indicates the mixing of the states. We find, that in particular the mixing between $\ket{x}$ and $\ket{xx}$ is strong. For example, at $t=0$ the state $\ket{\psi_2}$, having the middle energy, consists of \SI{26.4}{\percent} $\ket{g}$, \SI{23.2}{\percent} $\ket{x}$ and \SI{50.4}{\percent} $\ket{xx}$, i.e., the middle branch consists of $>\SI{50}{\percent}$ $\ket{xx}$ at the peak of the first pulse, while it corresponds to $\ket{x}$ before and after the pulse.

Having in mind that the goal is to excite either the exciton or the biexciton, it is desirable to find a second laser which induces transitions between the dressed states $\ket{\psi_{1,2,3}}$. 
%
\subsection{Driving the dressed states}
To gain more insight into the transitions between the dressed states, the Hamiltonian of the system is transformed to the dressed state basis. Similar to Eq.~\eqref{eq:interaction_dressed}, the interaction Hamiltonian $\tilde{H}_{\Omega_2}$ in the biexciton system not only results in additional energy contributions to the dressed states, that lead to the oscillatory behavior visible in Fig.~\ref{fig:dressed_state_diff_x}(a) and Fig.~\ref{fig:dressed_state_diff_b}(a), but also to a coupling of the dressed states. Here, we only want to consider the terms $\propto \ket{\psi_j}\bra{\psi_k}e^{-i\omega_\Delta t}$ with $\omega_\Delta>0$ for $j>k$ (i.e., $\ket{\psi_j}\bra{\psi_k}$ corresponds to a raising operator), which always leads to transitions driven by $\Delta_2<\Delta_1$.
\begin{align}
\begin{split}
    \omega_\Delta &= \Delta E_{jk}/\hbar \\
    \Rightarrow \Delta_2 &= \Delta_1 - \Delta E_{jk}/\hbar.
\end{split}
\end{align}
Here, $\Delta E_{jk}$ is the energy difference between two of the dressed states. 
In the transformed Hamiltonian $\tilde{H}_{\Omega_2}$, the coupling terms $\propto e^{-i\omega_\Delta t}$ of the second laser are
\begin{align}
\begin{split}
        \tilde{H}_{\Omega_2} =& -\frac{\hbar\Omega_2}{2}e^{-i\omega_\Delta t} \Big[\underbrace{(a_{2,g}a_{1,x}+a_{2,x}a_{1,xx})}_{:=\Omega_{12}(t)}\ket{\psi_2}\bra{\psi_1}\\
        &\underbrace{(a_{3,g}a_{1,x}+a_{3,x}a_{1,xx})}_{:=\Omega_{13}(t)}\ket{\psi_3}\bra{\psi_1}\\
        &\underbrace{(a_{3,g}a_{2,x}+a_{3,x}a_{2,xx})}_{:=\Omega_{23}(t)}\ket{\psi_3}\bra{\psi_2}\Big] + h.c. + ...
\end{split}\label{eq:biexciton_coupling_ds}
\end{align}
While in the bare-state picture the laser has the same coupling element for the $\ket{g}\leftrightarrow\ket{x}$ and $\ket{x}\leftrightarrow\ket{xx}$ transitions, in this picture there are different coupling terms $\Omega_{jk}$ for direct transitions between the dressed states $\ket{\psi_j}$ and $\ket{\psi_k}$.\\
\begin{figure}[t]
    \centering
    \includegraphics{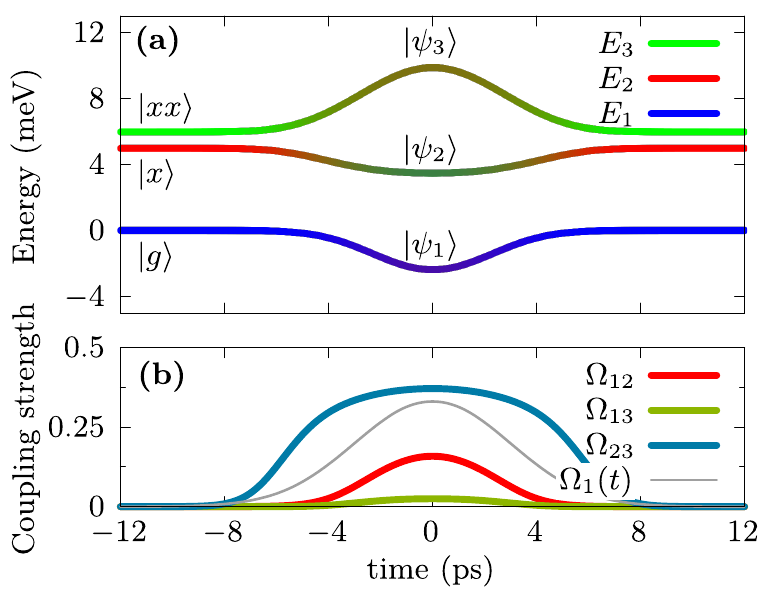}
    \caption{(a) Dressed-state energies $E_{1,2,3}$ for the biexciton system interacting with a single off-resonant pulse, $\sigma_1=\SI{3}{ps},\hbar\Delta_1=\SI{-5}{meV}, \alpha_1=27\pi$. The color indicates how the dressed states mix from the bare states. 
    (b) Coupling coefficients given in Eq.~\eqref{eq:biexciton_coupling_ds}, corresponding to the pulse in (a) as well as the laser envelope (arb. units). The time interval as well as the strength of the coupling is different for all three possible transitions, with the $\ket{\psi_2}\leftrightarrow\ket{\psi_3}$ transition showing by far the strongest coupling.}
    \label{fig:dressed_states_biexciton_coupling}
\end{figure}
Figure~\ref{fig:dressed_states_biexciton_coupling}(b) shows the time-dependent coupling coefficients given in Eq.~\eqref{eq:biexciton_coupling_ds} for the first pulse chosen in Fig.~\ref{fig:dressed_states_biexciton_coupling}(a). The couplings of the three possible transitions show significant differences, with $\Omega_{23}$ extending over the longest time period and having the largest amplitude. The transition coupled with $\Omega_{12}$ shows about half the coupling strength, while $\Omega_{13}$ only leads to a weak direct coupling of states $\ket{\psi_1}$ and $\ket{\psi_3}$.\\
Due to these simultaneous transitions between the dressed states, no clear criteria for high excitation of the exciton or biexciton state using a second pulse follows from this picture and we use a numerical parameter search.


\subsubsection{Exciton preparation}
\begin{figure}[t]
    \centering
    \includegraphics{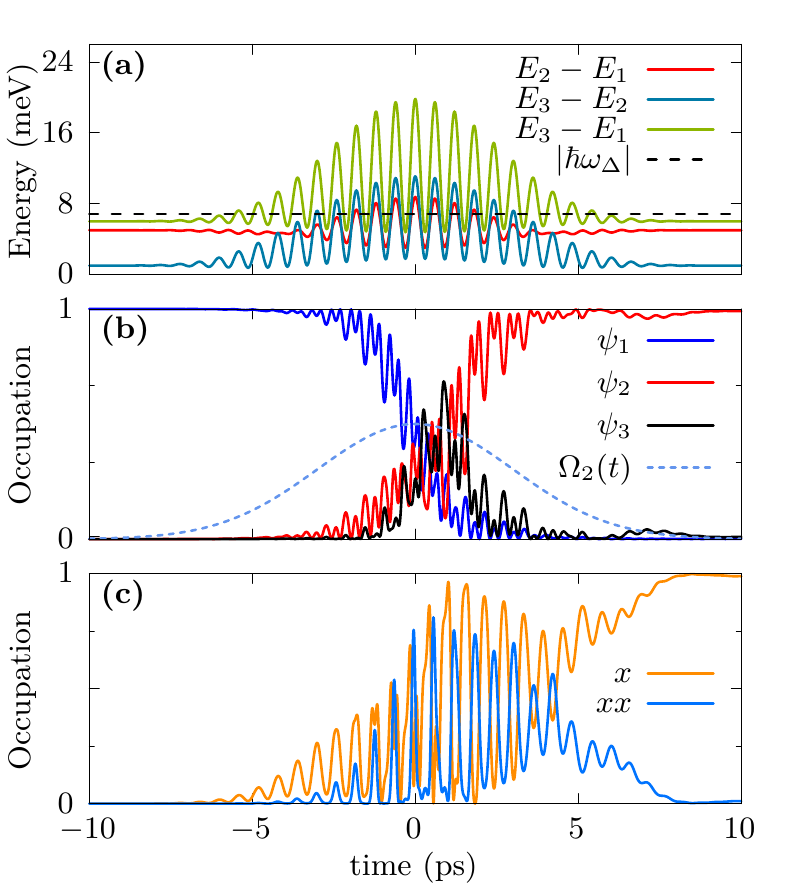}
    \caption{(a) Energy differences of the dressed states including the interaction Hamiltonian $\tilde{H}_{\Omega_2}$ containing a pulse with $\sigma_2=\SI{3}{ps},\hbar\Delta_2=\SI{-11.83}{meV},\alpha_2=22.8\pi$. The black dashed line depicts the energy of the second laser in this frame of reference. 
    (b) Dressed-state occupations. 
    The dashed blue line indicates the second pulse. (c) Occupation of exciton and biexciton.
    }
    \label{fig:dressed_state_diff_x}
\end{figure}
To excite the exciton, the dressed state $\ket{\psi_2}$ has to be populated by a second pulse, since this state evolves into $\ket{x}$ in the adiabatic undressing process. We use the same first pulse as in Fig.~\ref{fig:dressed_states_biexciton_coupling}(a). Numerical searching for parameters leads to a second pulse with $\sigma_2=\SI{3}{ps},\hbar\Delta_2=\SI{-11.83}{meV},\alpha_2=22.8\pi$. 
Figure~\ref{fig:dressed_state_diff_x}(a) shows the energy separation of the dressed states including the effect of the second pulse. The strong second laser leads to a high-amplitude oscillation of the dressed-state energies, such that the energy of the second laser (depicted as the black dashed line) overlaps with the energy differences of the dressed states in a prolonged time window. Panel (b) shows the dressed-state occupations. Similarly to the case of the two-level system, the occupation is transferred between the dressed states in an oscillatory fashion. However, the transfer starts only after the laser energy matches the splitting of the dressed states and can drive the transitions $\ket{\psi_1}\leftrightarrow\ket{\psi_2}$ and $\ket{\psi_2}\leftrightarrow\ket{\psi_3}$. 
While the energy also matches the $\ket{\psi_1}\leftrightarrow\ket{\psi_3}$ transition, this transition is strongly suppressed, as was shown in Fig.~\ref{fig:dressed_states_biexciton_coupling}(b). Nevertheless, $\ket{\psi_3}$ is populated during the dynamics due to the stronger $\ket{\psi_2}\leftrightarrow\ket{\psi_3}$ coupling. With the selected parameters, the occupation switches completely to the dressed state $\ket{\psi_2}$, which after the pulses corresponds to the exciton state. The bare state populations [panel (c)] show that during the pulses, both the exciton and biexciton state are addressed by the process, before the population completely switches to the exciton state.

\begin{figure}[t]
    \centering
    \includegraphics{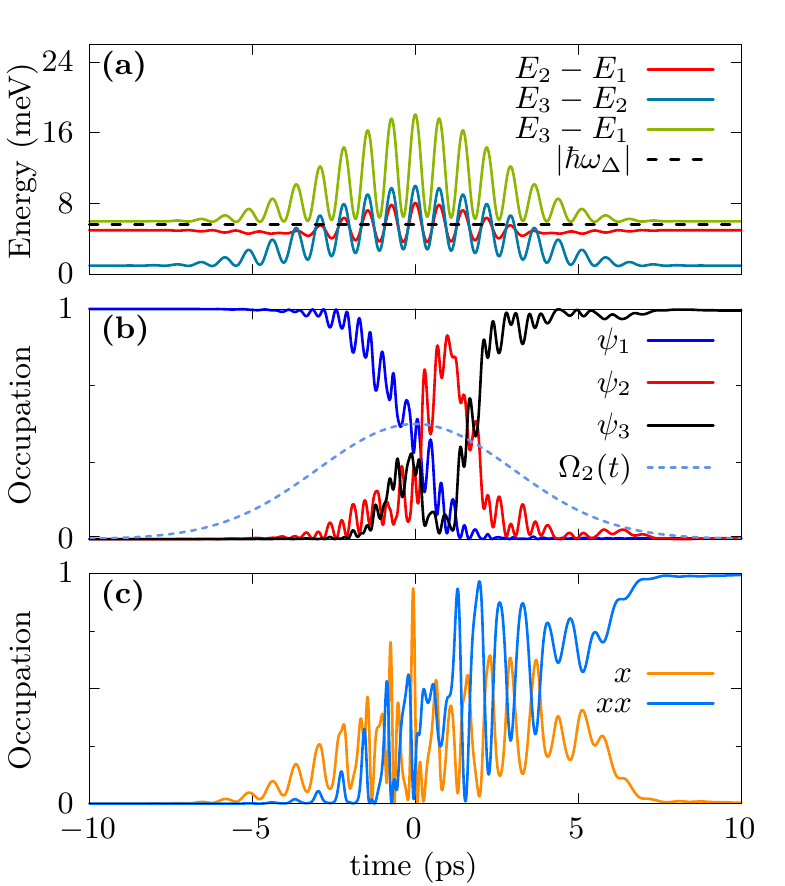}
    \caption{
    Same as Fig.~\ref{fig:dressed_state_diff_x}, but addressing the biexciton. For this, the detuning and pulse area of the second pulse are changed to $\hbar\Delta_2=\SI{-10.64}{meV},\alpha_2=17.49\pi$.
    }
    \label{fig:dressed_state_diff_b}
\end{figure}

\subsubsection{Biexciton preparation}
Next, we analyze the dressed-state dynamics for a second pulse that results in complete population transfer to the biexciton state. The first pulse is the same as used before, only the detuning and pulse parameters of the second pulse are changed to $\hbar\Delta_2=\SI{-10.64}{meV}$ and $\alpha_2=17.49\pi$. The dressed state analysis for this case is shown in Fig.~\ref{fig:dressed_state_diff_b}. Now, the second detuning is slightly lower than in the previous case, so in panel (a) the energy of the second laser (black dashed line) does not cross the energy difference $E_3-E_1$ anymore, but lies more centrally towards the energies needed for the $\ket{\psi_1}\leftrightarrow\ket{\psi_2}$ and $\ket{\psi_2}\leftrightarrow\ket{\psi_3}$ transitions. Looking at the dressed-state occupations in panel (b), the occupation is first transferred from $\ket{\psi_1}$ to $\ket{\psi_2}$, similar to the previous case. Subsequently, the occupation is fully transferred to dressed state $\ket{\psi_3}$. After the undressing resulting from smoothly switching off the pulse, this corresponds to the biexciton state. Accordingly, in Fig.~\ref{fig:dressed_state_diff_b}(c), we see that during the pulse again both exciton and biexciton are addressed, while after the pulse only the biexciton is populated.\\
As such, the scheme can also be used to off-resonantly excite the biexciton, which of course is also possible using two-photon resonant transitions \cite{schweickert2018ondemand}. However, it was shown recently that this hinders optimal entanglement-generation \cite{seidelmann2022two}, so other options to excite the biexciton can be beneficial in the future.

\section{Conclusion}
In this work we analyzed the impact of two simultaneous laser pulses on a quantum emitter, that are detuned to the transition energy. Off-resonant pulse pairs are used in two-color excitation like the SUPER scheme as proposed theoretically \cite{bracht21swingup} and implemented for quantum dots \cite{karli2022super} to populate the excited state in a two-level system. Using the dressed-state picture, we find that choosing one pulse uniquely determines the optimal parameters for the other pulse. For the two-level system, we provide analytic formulas [see for example Eq.~\eqref{eq:energy2_below} together with Eq.~\eqref{eq:pulsearea}] for the second pulse for optimal preparation of the excited state. We have demonstrated, that many combinations of positive and negative detunings can be used to obtain the SUPER effect, leading to flexibility in the design of future experiments. Further, we discussed that in a three-level system, the two-pulse excitation can be used to selectively excite the exciton or biexciton. These results enable the operation of single-photon or entangled photon sources with the help of the SUPER excitation scheme.
\bibliography{bibfile}

\end{document}